\documentclass[reprint,superscriptaddress,showpacs, amsmath,amssymb]{revtex4-1}
\usepackage{graphicx} 

\usepackage{amsfonts}
\usepackage{amssymb,color}

\begin{document}
\author{M.V. Hakobyan}
\affiliation{Yerevan State University, 1 Alex Manookian, Yerevan 0025, Armenia}
\affiliation{Institute for Physical Research, NAS of Armenia, Ashtarak 0203, Armenia}
\author{V.M. Red'kov}
\affiliation{Institute of Physics of NAS of Belarus, F. Skarina Avenue 68, Minsk 220072, Belarus}
\author{A.M. Ishkhanyan}
\affiliation{Institute for Physical Research, NAS of Armenia, Ashtarak 0203, Armenia}

\title{Adiabatic asymmetric scattering of atoms in the field of a standing wave }

\begin{abstract}
A model of the asymmetric coherent scattering process (caused by initial atomic wave-packet splitting in the momentum space) taking place at the large detuning and adiabatic course of interaction for an effective two-state system interacting with a standing wave of laser radiation is discussed. We show that the same form of initial wave-packet splitting may lead to different, in general, diffraction patterns for opposite, adiabatic and resonant, regimes of the standing-wave scattering. We show that the scattering of the Gaussian wave packet in the adiabatic case presents {\em refraction} (a limiting form of the asymmetric scattering) in contrast to the {\em bi-refringence} (the limiting case of the high-order narrowed scattering) occurring in the resonant scattering.
\\
\\ \textbf{PACS number(s):} 37.10.Vz Mechanical effects of light on atoms, molecules, and ions, 03.75.-b Matter waves, 37.25.+k Atom interferometry techniques, 03.75.Dg Atom and neutron interferometry, 03.75.Be Atom and neutron optics
\\

\end{abstract}

\maketitle

\subsection{ Introduction}

The observation of asymmetric diffraction in experiments involving the scattering of sodium atoms by a field of two short counterpropagating pulses of laser radiation \cite{Grinchuk1}, \cite{Grinchuk2} has stimulated several developments \cite{Romanenko}-\cite{VitkovskyPrants} intended to explore these peculiarities of the Kapitza-Dirac diffraction \cite{Kapitza-Dirac},\cite{Gould} for applications in atom interferometry \cite{Berman}, \cite{Cronin} and atom lithography \cite{AtomLithography}, \cite{MeschedeMetcalf} using atom optics techniques \cite{Adams}, \cite{Meystre}. These efforts have led to advanced representations on the scattering of atoms by standing waves extending the diversity of the scenarios of interference occurring during the interaction of atoms with the field of optical lattices and, in general, mechanical action of light on the matter waves \cite{CookBernhardt}-\cite{Kazantsev}.

The asymmetric scattering model employs secondary quantum-mechanical interference during interaction with the radiation field to achieve different intended target states. This interference is due to superposition initial states. It has been shown that the preparation of particles prior to interaction in specific (in general, opto-mechanically mixed) states is able to dramatically alter the interaction pattern \cite{Romanenko}-\cite{Ishkhanyan1}. A basic example of such a change is the strong asymmetry in the scattering pattern in the case when the atomic wave packet is initially split into two momentum peaks differing by an odd number of photon momenta \cite{Ishkhanyan1}. Even more advanced are the various elaborate initial superposition states \cite{Romanenko}-\cite{PazgalevRozhdestvenskii} that may result in a large amplitude coherent accumulation of the momentum on the internal energy levels caused by: single photon exchange \cite{Muradyan1}, narrowing of the interference fringes of the diffraction pattern \cite{Ishkhanyan2}, standing-wave refraction of atoms with initial Gaussian distribution of amplitudes by momenta \cite{Ishkhanyan3}, etc. These effects suggest more flexibility in the control of atomic motion and hence can be useful in atom optics, in particular, in atom interferometric and atom nanolithographic applications (see, e.g., \cite{Muradyan2}).

The peculiarities of asymmetric scattering are expressed further when dealing with the close neighborhood of exact resonance or when the fast switching on/off of a laser pulse is involved in the process. This is because, in these cases, stronger excitation of the system is achieved. Besides, the sudden inclusion of the interaction, an essentially non-adiabatic process, suggests more flexibility in choosing different preparation states. However, it is understood that many of the explored effects can also be observed in the adiabatic regime, i.e. at large detunings of the wave frequency and the slow course of the interaction. Since the adiabatic interaction schemes as a rule suggest more robust technologies, complementary discussions of the adiabatic model of the asymmetric scattering are in demand to clarify the potential for controlling the diffraction picture of atomic wave-packets in this regime.

In the current paper, we first present the simplest scheme for the coherent diffraction of atoms by a standing wave occuring strongly
asymmetrically in the adiabatic regime - the case where the initial atomic wave packet involves only two translation states whose momenta differ by two photon momenta. We then reveal that the same form of initial coherent superposition state of the atom may cause qualitatively different effects for adiabatic and resonant standing-wave scattering (though this is not necessarily the case for all initial wave-packets). Finally, we demonstrate that the evolution of the Gaussian momentum distribution of amplitudes at adiabatic standing-wave scattering presents a strongly asymmetric, with respect to the initial momentum direction, diffraction (with minor deformation of the wave packet form) while at resonance the same wave packet undergoes symmetric scattering (analogous to the high order narrowed diffraction discussed in \cite{Ishkhanyan2}).

It should be noted that the scattering model in the form presented below covers only some of the important features of the asymmetric diffraction. For instance, it does not describe the oscillatory dependence of the scattering amplitude on the resonance detuning \cite{Grinchuk2}. However, the inspection of the known developments for the resonant case shows that it is possible to include these peculiarities by means of modification of the intermediate preparation states and taking into account the first-order non-adiabatic corrections.

\subsection{Adiabatic asymmetric scattering}

The dynamics of an effective two-state system in the field of a standing wave $E=2E_0f(t)\cos (kz)\cos (\omega t)$ with a slowly varying envelope $f(t)$ at small interaction times (in the absence of spontaneous emission) is described, in the rotating wave and Raman-Nath approximations, by the time dependent Schr\"{o}dinger equations for the probability amplitudes of the states $a_{1,2}$
\begin{eqnarray}
i\frac{da_1}{dt} &=&2U_0^{*}f(t)\cos (kz)e^{-i\Delta t} a_2,  \label{1a}
\\
i\frac{da_2}{dt} &=&2U_0f(t)\cos (kz)e^{+i\Delta t} a_1,
\label{1b}
\end{eqnarray}
where $\Delta =\omega _{21}-\omega _0$ is the resonance detuning, $%
U_0=-dE_0/(2\hbar )$ is the peak Rabi frequency of the travelling wave, $d$ is the dipole moment of the transition under consideration.

In the adiabatic regime of large detuning, slow inclusion and variation of the interaction, $\left| \Delta \right| t \gg 1$, this system is reduced, via adiabatic elimination of the excited state (see, e.g., \cite{Dulcie},\cite{Martin}), to a simple first order equation
\begin{equation}
i\frac{da_1}{dt}=-\frac{4\left| U_0\right| ^2}\Delta f^2(t) \cos
^2(kz) \; a_1,
\label{3}
\end{equation}
the solution of which is straightforward:
\begin{equation}
a_1(t)=a_1(0) e^{i\frac{2\left| U_0\right| ^2}\Delta \tau } e^{i
\frac{2\left| U_0\right| ^2}\Delta \tau \cos 2kz},
\label{4}
\end{equation}
where $\tau $ is the integral of the square of the field envelope:
\begin{equation}
\tau =\int\limits_0^tf^2(t)dt.
\label{5}
\end{equation}

To present the simplest model for the asymmetric scattering in the adiabatic regime, we consider, following \cite{Ishkhanyan2}, the scattering caused by the initial conditions of the form
\begin{eqnarray}
a_1(0) &=&\newline
\sum\limits_{m=-\infty }^{+\infty }\alpha _{2m}e^{i2mkz} \varphi (z),
\nonumber \\
a_2(0) &=&0,  \label{6}
\end{eqnarray}
which, evidently, can be created using adiabatic processes. The
solution of the coherent diffraction problem in the momentum representation is (for simplicity, we suppose that before the preparation in the state (\ref{6}) the atom had an exactly defined momentum $p_0$, that is $\int \varphi (z)e^{-ikz/\hbar }d(z)=\delta (p-p_0)$):
\begin{eqnarray}
a_1 =
e^{i\frac{2\left| U_0\right| ^2\tau }\Delta } \sum\limits_n (\textrm{sign}\Delta \cdot i)^{n/2}{1+(-1)^n \over 2} \cdot
\nonumber \\
{\left( \sum\limits_m(\textrm{sign}\Delta \cdot
i)^{-m}\alpha _{2m}J_{{n\over2}-m}\left( \frac{2\left| U_0\right| ^2\tau }{\left|
\Delta \right| }\right) \right) \delta (p-p_0-n\hbar k)},
\label{7}
\end{eqnarray}
where $J$ is the Bessel function. The corresponding probability of absorbing $n$ photons is written as:
\begin{equation}
W_n(t)=\frac{1+(-1)^n}2\left| \sum\limits_m(\textrm{sign}\Delta \cdot i)^{-m}\alpha
_{2m}J_{n/2-m}\left( \frac{2\left| U_0\right| ^2\tau }{\left| \Delta \right|
}\right) \right| ^2.
\label{8}
\end{equation}

If the initial wave packet is not split in the momentum space, i.e. if $\alpha _{2m}=0$ at $m\neq 0$, then the solution (\ref{7}) becomes the well-known expression \cite{Gould},\cite{CookBernhardt}:
\begin{equation}
W_n(t)=\frac{1+(-1)^n}2J_{n/2}^2(u),\quad u=\frac{2\left| U_0\right| ^2\tau }{\left| \Delta \right| },
\label{9}
\end{equation}
that describes symmetric, with respect to the initial atomic momentum $p_0$, diffraction pattern.

However, as it is readily seen from Eq. (\ref{8}), the situation is qualitatively changed at the initial conditions of splitting. It is seen that a secondary interference takes place; during diffraction by the standing wave the diffraction peak sets \{$J_{n/2-m}$\}, originated from the corresponding peaks of the initial wave packet, overlap. This interference significantly changes the scattering pattern. Indeed, consider, for instance, the simplest case when the initial wave packet is split into only two peaks:
\begin{eqnarray}
a_1(0) &=&\newline
(\alpha _0+\alpha _2e^{i2kz})\; \varphi (z),  \nonumber \\
a_2(0) &=&0.  \label{10}
\end{eqnarray}
For the probability of the $n$-th diffraction order we then get a strongly asymmetric scattering:
\begin{equation}
W_n(t)=\frac{1+(-1)^n}2\left| \alpha _0J_{n/2}(u)+\frac{\alpha _2}{i\;
\textrm{sign}\Delta }J_{n/2-1}(u)\right| ^2.
\label{11}
\end{equation}
Indeed, taking into account the equality $J_{-n}=(-1)^nJ_n$, we have:
\begin{equation}
W_n(t)=\frac{1+(-1)^n}2\left\{
\begin{array}{c}
\left| \alpha _0\right| ^2J_{n/2}^2+\left| \alpha _2\right| ^2J_{n/2-1}^2- \\
\frac{2 \mathop{\rm Im}
(\alpha _0\alpha _2^{*})}{\textrm{sign}\Delta }J_{n/2}J_{n/2-1},\text{ }n>0, \\
\left| \alpha _0\right| ^2J_{-n/2}^2+\left| \alpha _2\right| ^2J_{-n/2+1}^2+ \\
\frac{2 \mathop{\rm Im}
(\alpha _0\alpha _2^{*})}{\textrm{sign}\Delta }J_{-n/2}J_{-n/2+1},\text{ }n<0.
\end{array}
\right.  \label{12}
\end{equation}

We note that the scattering probability (\ref{11}) has the same structure as the corresponding non-adiabatic probability for the exact resonance case with preliminary excitation of the atom by a travelling wave \cite{Ishkhanyan1}. Hence, the peculiarities of the diffraction process in the adiabatic and non-adiabatic regimes are qualitatively the same.

\begin{figure}
\centering
\includegraphics[width=260pt]{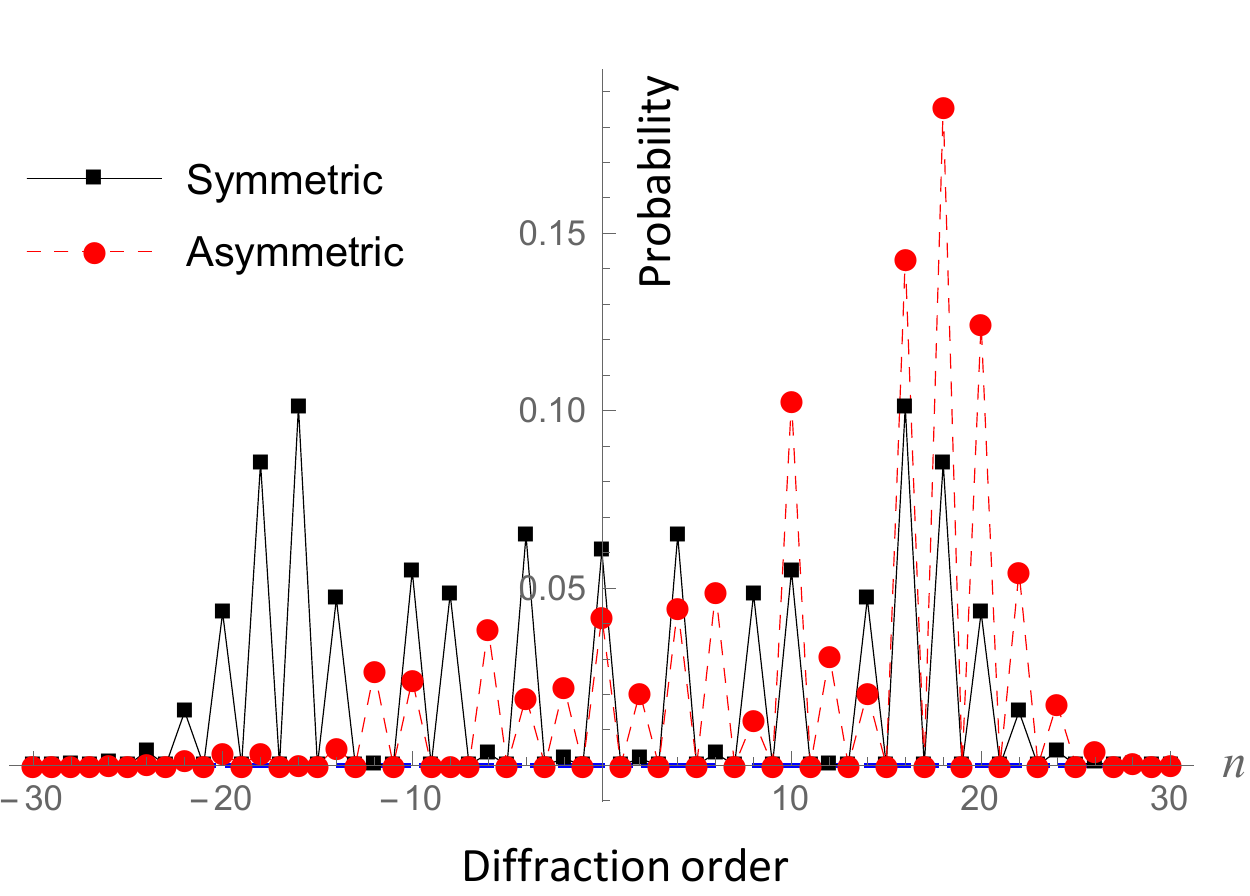}
\caption{Diffraction patterns at adiabatic interaction of atoms with a standing wave: (a) usual deflection - Eq. (\ref{9}), (b) asymmetric deflection - Eq. (\ref{11}), $\alpha _0=1/\sqrt{2},$ $\alpha _2=-i/\sqrt{2},$ $\Delta \tau=-500,U\tau=50.$}
\label{Fig.1}
\end{figure}

The scattering pattern asymmetry is defined as
\begin{eqnarray}
\Delta W(t) &=&\sum\limits_{n=1}^{+\infty }(W_{+n}-W_{-n})
\nonumber \\
&=&\left| \alpha
_2\right| ^2(J_0^2+J_1^2)-\frac{2\;{\rm Im}
(\alpha _0\alpha _2^{*})}{\textrm{sign}\Delta }(C_0-J_0J_1),
\nonumber \\
C_0 &=&\int\limits_0^u(J_0^2(u)+J_1^2(u))du,\quad C_0\left| _{u\rightarrow
\infty }\right. \approx 0.638.
\label{13}
\end{eqnarray}
Consequently, the maximum possible asymmetry is achieved at
\begin{equation}
{\rm Im}(\alpha _0\alpha _2^{*})=\pm 1/2
\label{14}
\end{equation}
($\left| \alpha _0\right| =\left| \alpha _2\right| =1/\sqrt{2}$); and it is the same as in the non-adiabatic case: at $u\rightarrow \infty $, more than 80\% of atoms deflects to a definite direction. However, the preferable direction of the deflection is determined, as it is seen from Eq. (\ref{12}), not only by the sign of ${\rm Im}(\alpha _0\alpha _2^{*})$, but also by the sign of the detuning. The total acquired momentum of the atom after the interaction is
\begin{equation}
\left\langle p\right\rangle =\sum\limits_{-\infty }^{+\infty }2n\hbar
k W_{2n}=2\hbar k\left( \left| \alpha _2\right| ^2-{\rm Im}
(\alpha _0\alpha _2^{*}) \frac{2\left| U_0\right| ^2\tau }\Delta
\right).
\label{15}
\end{equation}
The first term in the right-hand side of this equation, which is the momentum shift coming from the splitting of the initial state (\ref{6}), is not more than two photon momenta, but the second term (the result of the standing wave action) is not restricted. However, we note that the mean momentum increase is determined, in addition to the factor $\mathop{\rm Im}(\alpha _0\alpha _2^{*})$ coming from the initial conditions, by the parameter $u={2\left| U_0\right| ^2\tau } / {\left| \Delta \right|}$ which is supposed to be not too large within the Raman-Nath approximation. The approximation is valid if the gained kinetic energy of the atom $\epsilon_{kin}=(n \hbar k)^2/(2m)$  is small compared with the interaction energy $\epsilon=\hbar U_0$ for all diffraction orders  $n$. It follows from the properties of the involved Bessel functions that the maximum populated diffraction order is approximately $n_{max}\approx 2u$. Hence, should be $4u^2 \omega_{rec} \ll \min \{U_0,1/t\}$, where $\omega_{rec}=\hbar k^2/(2m)$ is the recoil frequency and $t$ is the interaction time. The diffraction pattern caused by the initial splitting (\ref{10}) at the maximum possible asymmetry is presented in Fig.\ref{Fig.1}. For the chosen parameters $U_0$ and $\Delta$ this pattern is consistent with the Raman-Nath approximation, e.g. for sodium atoms with $m=23$ amu and optical field with $\lambda \approx 0.5$ $\mu$m, for interaction times $\tau<10^{-7}$ s.

Thus, we have seen that the adiabatic interaction with the standing wave can also occur asymmetrically provided the initial wave packet is split in a special manner. It should be noted that, interestingly, the wave-packet (\ref{10}) has the same form as the one considered in \cite{Ishkhanyan2} (Eq.(21)) when discussing the narrowing of the interference fringes of the diffraction pattern at the exact resonance. Thus, the same form of initial wave-packet splitting may, in general, lead to different diffraction patterns for adiabatic and resonant standing-wave scattering.

Nevertheless, the wave packet (\ref{10}) itself can not be used to achieve both resonant narrowed (at $\left| \Delta \tau \right| \ll 1$) and adiabatic asymmetric (at $\left| \Delta \tau \right| \gg 1$) scatterings. The narrowing condition $\alpha_2=-\alpha _0$ \cite{Ishkhanyan2} and the condition for maximum asymmetry (\ref{14}) are not compatible since the phase conditions that should be imposed on $\alpha _0$ and $\alpha _2$ contradict. However, this incompatibility is not necessarily the case for all the possible wave-packets. We demonstrate this below by examining the behavior of the Gaussian wave packets. We will see that the adiabatic standing-wave scattering of a Gaussian wave packet presents {\em refraction} (a limiting form of asymmetric scattering, see also \cite{Ishkhanyan3}) in contrast to the {\em bi-refringence} (the limiting case of high-order narrowed scattering, \cite{Ishkhanyan2}) occurring with the same wave-packet at the resonant scattering.

Regarding the preparation of atoms in the states (\ref{6}), first we note that such distributions can be achieved, in general, in effective two-state systems when the terms $\alpha_{2m}e^{i2mkz}\varphi (z)$ correspond to the different levels that compose the effective ground state. Such a situation is the case, e.g. in the two-level systems with magnetic sublevels (see an example of such a preparation of atoms by a single elliptically polarized travelling wave pulse in \cite{IshkhanyanSublevels}). Alternatively, this distribution can be viewed as one corresponding to the same internal state of an atom. For instance, a combination of adiabatic rapid passage and multiphoton Bragg diffraction can be used to efficiently transfer many photon recoils of momentum $\hbar k$ to cold-atoms, thus creating two- or multi-peak distributions of the needed structure \cite{Kasevich}. Creation of a Gaussian momentum profile starting from the atoms prepared in a very narrow distribution around $p=0$, then adiabatically ramping on a one-dimensional optical lattice and further suddenly spatially shifting the lattice by 1/4 of the lattice period, is reported in \cite{Raizen}.

Further, as a systematic method to create the desired distributions of the populations in the multi-level systems one may apply different STIRAP schemes \cite{BergmannTheuerShore} involving combinations of traveling and standing waves. An example of such a process leading to the creation of discrete wave-packets of exponential distribution by momenta is suggested in \cite{Ishkhanyan2}. Momentum state preparation of a two-level atom using two-stage Kapitza-Dirac diffraction of an initially single-momentum atomic beam was recently discussed in \cite{A.Movsisyan-G.Muradyan}. Gaussian wave packets can be prepared by exciting atoms to Rydberg states \cite{AlberZoller} (for preparation of molecular wave packets by femtosecond pulse technology see, for instance, \cite{Gruebele}).

\subsection{Diffraction in the case of initial Gaussian distribution by momenta}

\begin{figure}
\centering
\includegraphics[width=260pt]{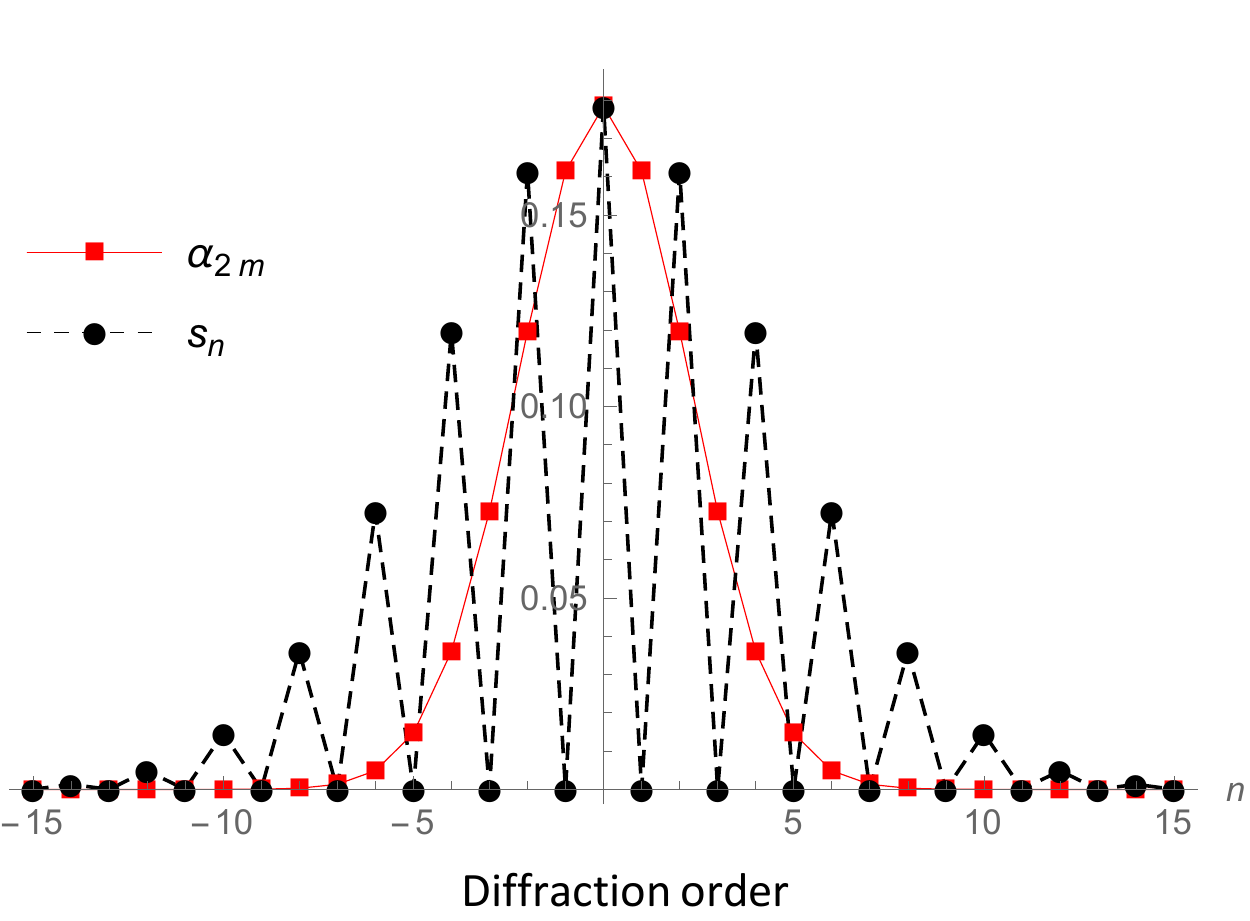}

\caption{Representations of the same Gaussian initial wave packet ($M=10$)
for adiabatic ($\alpha_{2m}$) and resonant($s_{m}$) scattering regimes.}
\label{Fig.2}
\end{figure}

For a discrete Gaussian wave-packet involving only even orders (see Fig.\ref{Fig.2}):
\begin{equation}
\alpha _{2m}(m)={\frac {e^{i(\alpha +(\textrm{sign}\Delta )\pi /2)m}} {(\pi M)^{1/4}}} e^{-{\frac{m^2}{2M}}},
\label{16}
\end{equation}
where $M$ is the distribution half-width, the adiabatic scattering probability (\ref{8}) is rewritten as ($\nu=n/2-m$)
\begin{eqnarray}
W_n &=\frac {1+(-1)^n}{2}\left| \frac{e^{i\alpha n/2}}{(\pi M)^{1/4}}  \sum\limits_{\nu=-\infty
}^{+\infty }e^{-i\alpha \nu -(n/2-\nu )^2/(2M)} {J_\nu(u)} \right| ^2
\nonumber
\\
&\equiv \frac {1+(-1)^n}{2} \left| e^{i\alpha
n/2}I_{n/2}(u)\right| ^2,
\label{17}
\end{eqnarray}
where $u=2\left| U_0\right| ^2\tau /\left| \Delta \right|$. The
behavior of the functions $I_n(u)$ was studied in \cite{Ishkhanyan3}.
The approach employs the following exact linear differential-difference equation:
\begin{equation}
\frac{dI_n}{dn}=-\frac nMI_n+\frac u{2M} \left(e^{i\alpha }I_{n+1}+e^{-i\alpha}I_{n-1} \right),
\label{18}
\end{equation}
which is readily derived by differentiating $I_n$ and using the identity $2 \nu J_{\nu}(u)=u (J_{\nu+1}+J_{\nu-1})$.

Examining the solution of this equation, we note that at the beginning of the scattering, $u=0$, the solution is, as expected, the Gaussian
\begin{equation}
I_n= c_{0} e^{-\frac{n^2}{2M}}
\label{19}
\end{equation}
with $c_{0}$ being the pre-factor of the exponent in Eq. (\ref{16}). Furthermore, dividing the equation by $n/M$ and passing to the variable $l=n^2/(2M)$ ($\Leftrightarrow dl=(n/M)dn$) we see that the term in the brackets in Eq. (\ref{18}) becomes proportional to $u/(2n)$, hence, it can be neglected for large diffraction orders such that $\left|n\right| \gg u$. Thus, the dynamics of the wave packed given by $I_{n/2}(u)$ is effectively localized within the interval $-2u \le n \le 2n$. We will see that during that time the wave-packet moves as a whole within this interval and slightly changes its form.

To discuss the diffraction details, we expand $I_{n\pm 1}$ into the Taylor series at the point $n$ and keep the first three terms. The resultant equation reads
\begin{equation}
\frac{dI_n}{dn}=-\frac{n}{M} I_n+\frac{u}{M} \left( (\cos \alpha )\left[ I_n+\frac{1}{2}\frac{d^2I_n}{dn^2}\right] +i(\sin \alpha )\frac{dI_n}{dn}\right).
\label{20}
\end{equation}

If $\cos \alpha=0$, the equation is reduced to a first-order one, the solution of which is again given by a Gaussian:
\begin{equation}
I_n= c_{0} e^{-\frac{n^2}{2(M \pm iu)}},
\label{21}
\end{equation}
where now $c_{0}=\left[\pi M \left(1+{u^2}/{M^2} \right)\right]^{-1/4}$. The distribution half-width is $\sqrt{M^2+u^2}$, hence, the wave-packet steadily broadens during the time.

If $\cos \alpha \not=0$, equation (\ref{20}) is reduced to the Airy equation \cite{Ishkhanyan3}. Accordingly, the solution of the diffraction problem finite at $n\rightarrow \pm \infty $ in this case is written in terms of the Airy function of the first kind \cite{Abramowitz}. However, before discussing this solution, it is helpful to take a look at the solution derived if one only neglects the second-derivative term in Eq. (\ref{20}):
\begin{equation}
I_n= c_{0} e^{-\frac{(n-(\cos \alpha)u)^2}{2(M + i (\sin \alpha) u)}} ,
\label{22}
\end{equation}
where now $c_{0}=\left[\pi M \left(1+{(\sin \alpha)^2 u^2}/{M^2} \right)\right]^{-1/4}$. This solution indicates that in the case of non-zero $\cos \alpha$ the wave-packet in general broadens and moves as a whole in the momentum space, the broadening being defined by $\sin \alpha$ and the displacement being proportional to $\cos \alpha$.

Consider now the exact solution of Eq. (\ref{20}):
\begin{equation}
I_n=c_{0}e^{hN}Ai(N+h^2),
\label{23}
\end{equation}
where the parameters $N$, $h$ are written as
\begin{equation}
N= \frac {n-(\cos \alpha )u}{2^{-1/3}\left(\left|\cos \alpha \right|u\right) ^{1/3}}, \quad h=\frac{M-i(\sin \alpha ) u}{2^{1/3}\left(\left|\cos \alpha \right|u\right)^{2/3}},
\label{24}
\end{equation}
and the constant $c_{0}$ is defined from the normalization condition. This is a localized wave-packet the properties of which are very controlled by the imaginary part of the argument $z=N+h^2$ of the Airy function: $\operatorname{Im}(z)=\operatorname{Im}(h^2) \sim \sin (\alpha)$. Indeed, consider the case $\sin \alpha=0$ when the argument is real. Then, $\arg z=0$ for $n>u$ and $\arg z=-\pi$ for $n<u$. The asymptotes of the Airy function for large real argument $z=x$ are known to be $Ai(x)\sim e^{-(2/3)x^{3/2}}/ x^{1/4}$ if $x>1$ and $Ai(x)\sim {\sin (x+\pi/4)}/ x^{1/4}$ if $x<-1$ \cite{Abramowitz}. Hence, the localization at the packet side for which $N<0$ is due to the factor $e^{hN}$ (since $\operatorname{Re}(h)$ is positive) and the wave packet is localized at its other side due to the Airy function asymptote.

\begin{figure}
\centering
\includegraphics[width=260pt]{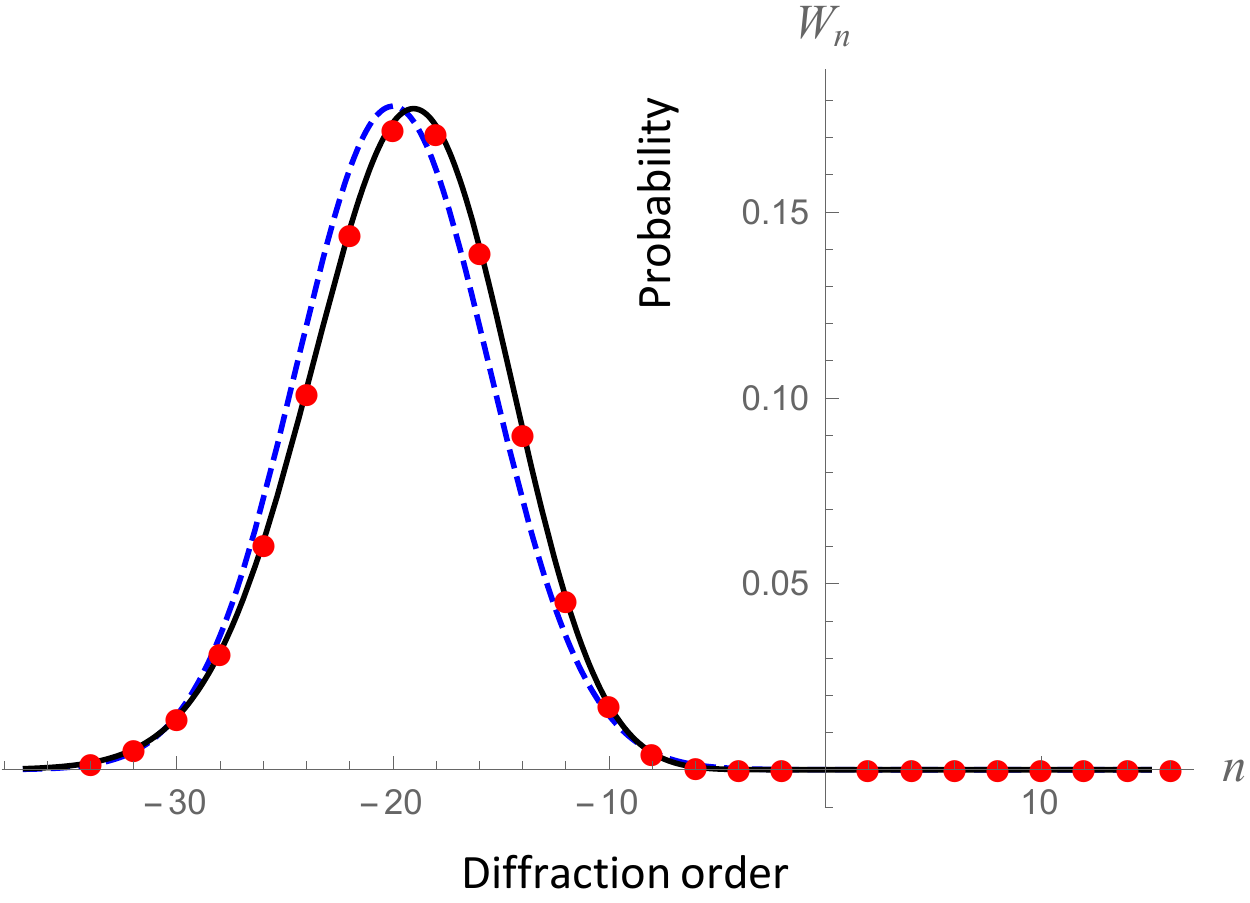}

\caption{Adiabatic diffraction pattern of the Gaussian wave packet with $M=10, \alpha=\pi$: $\Delta \tau=500, U\tau=50$. Dashed line - moving Gaussian approximation Eq.(22). Solid line - Airy function solution Eqs. (23),(24).}
\label{Fig.3}
\end{figure}

If $\sin \alpha \not=0$, then $\arg z \not=0, \pi$ for any $n$ (because $N$ is real and $\operatorname{Im}(h)\not=0$). The real part of the argument of the Airy function becomes zero near the point $n\simeq (\cos \alpha) u + \textrm{O} (1/n)$. Going far away from this point, that is at $(n-(\cos \alpha)u)\rightarrow \pm\infty$, depending on the sign of $(\sin \alpha)/n$, $\arg z$ either tends to zero or $-\pi$. The asymptote of the Airy function for large $\left| z \right|$ is $Ai(z)\sim e^{-(2/3)z^{3/2}}/ z^{1/4}$ in the sector $\left|\arg z \right| < \pi$ and $Ai(z)\sim 1/ z^{1/4}$ in the sector $\left| \pi- \arg(z) \right| < \epsilon \ll 1$.
Accordingly, for the side of the distribution for which $(\sin \alpha)/n$ is positive ($\operatorname{Im}(z)<0$) the localization is due to the factor $e^{hN}$ (the Airy function adds negligibly small oscillations) and the localization is due to the asymptote of the Airy function in the opposite side of the packet. We conclude by noting that the Airy-function solution (\ref{23})-(\ref{24}) presents a highly accurate approximation. The comparison with the exact numerical result and the moving Gaussian approximation (\ref{22}), which provides qualitatively rather a good description, is shown on Fig.\ref{Fig.3}).

Thus, during the time evolution, the wave packet always remains localized moving in the momentum space if $\cos a\neq 0$ and broadening if $\sin a\neq 0$. If $\alpha =\pi/2\pm \pi k,k=0,1,2...$ the distribution peak does not move; it just broadens. In contrast, if $\alpha =\pm \pi k,k=0,1,2...$ the broadening is absent in the first approximation and the distribution displaces as a whole (see Fig.\ref{Fig.3}). The displacement of the peak position as well as the distribution broadening are determined by the parameter $u=2\left| U_0\right| ^2\tau /\left| \Delta \right|$. Thus, the adiabatic scattering of the Gaussian wave-packet (\ref{16}) presents, approximately, a refraction to a definite angle controlled by the interaction time with the light field. This picture significantly differs from that for the resonant scattering regime.

Indeed, consider the resonant diffraction of the same Gaussian wave packet (\ref{16}). The scattering probability $W_n^{r}$ in this case is written as \cite{Ishkhanyan3}
\begin{equation}
W_n^{r}(t)=\left| \sum\limits_mi^ms_mJ_{n-m}(2Ut)\right| ^2,  \label{25}
\end{equation}
where $t$ is the interaction time and the corresponding initial-state vector $s_m$ for a distribution of the form (\ref{6}) is defined as
\begin{equation}
s_m(m)=\frac{1+(-1)^m}2\alpha _{2m}(m/2).
\label{26}
\end{equation}
(see Fig. 2).

\begin{figure}
\centering
\includegraphics[width=260pt]{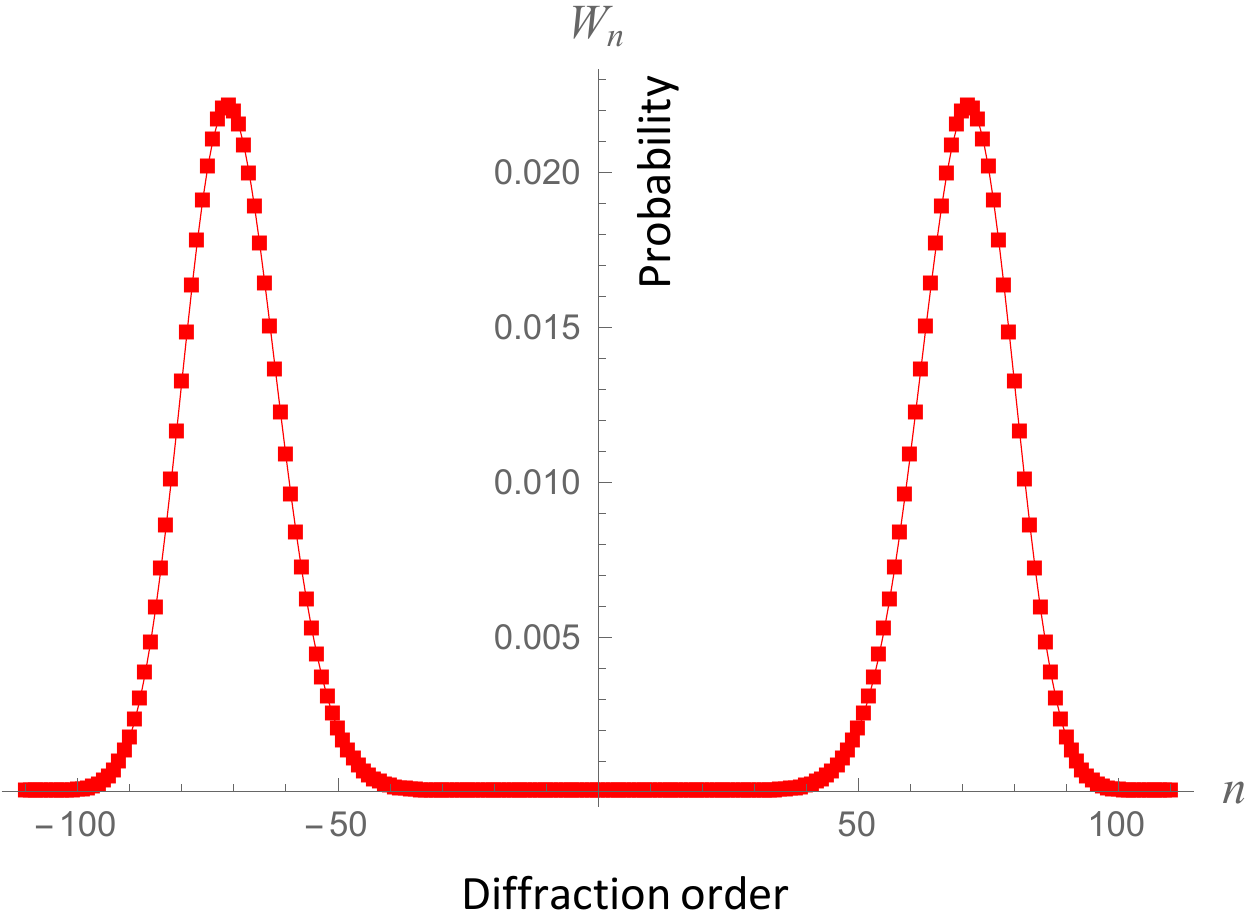}

\caption{Resonant diffraction pattern of the Gaussian wave packet (\ref{16}) with $M=10, \alpha=\pi$: $\Delta =0,Ut=50$.}
\label{Fig.4}
\end{figure}

Rewriting $i^m s_m$ as ($m \rightarrow n-\nu$)
\begin{equation}
i^m s_m=\frac{e^{-\frac{(n-\nu )^2}{2 M_1}+i \beta  (n-\nu )}+(-1)^{n-\nu } e^{-\frac{(n-\nu )^2}{2 M_1}+i \beta  (n-\nu )}}{\sqrt{2} \sqrt[4]{\pi  M_1}}
\label{27}
\end{equation}
with $M_1=4 M$ and
\begin{equation}
\beta = \frac{1}{4} (2\alpha +2\pi +\pi \; \text{sign}(\Delta )),
\label{28}
\end{equation}
thus splitting the sum in Eq. (\ref{25}) into two parts, then passing to a new summation variable $\nu=n-m$ and further using the identity $(-1)^\nu J_{\nu}(u)=J_{\nu}(-u)$, we get
\begin{equation}
W_n^{r}=\frac 1{2\sqrt{\pi }}\left|
I_n(u_{r})+(-1)^nI_n(-u_{r})\right| ^2 ,
\label{29}
\end{equation}
where $u_{r}=2Ut$ and
\begin{eqnarray}
I_n^{r}(u_{r}) = \sum\limits_{-\infty
}^{+\infty} {e^{-i\beta \nu -(n-\nu )^2/2M_1}\cdot \frac 1{M^{1/4}}J_\nu (u_{r})}.
\label{30}
\end{eqnarray}
As it is immediately seen, equation (\ref{29}) describes a symmetric two-fringe scattering pattern. Hence, the resonant scattering of the Gaussian wave packet (\ref{16}) presents bi-refringence. This is demonstrated in Fig.\ref{Fig.4}, where the presented graph has been calculated using the exact equation (\ref{25}).

We note that $I_n^{r}(u_{r})$ given by Eq.(\ref{30}) is exactly the same function as $I_n(u)$ (with parameters altered as $u \rightarrow u_r$, $M \rightarrow M_{1}/4$, and $\alpha \rightarrow \beta$ according to Eq. (\ref{28})) that we used in treating the adiabatic scattering. Thus, one may use the above moving Gaussian approximation (\ref{22}) or the Airy-function solution (\ref{23})-(\ref{24}) to accurately explore the interference of the fringes at resonant scattering. At $\alpha =\pm \pi k$, $k=0,1,2...$ and large enough $u_r$ when the functions $I_n(u_{r})$ and $I_n(-u_{r})$ practically do not overlap (i.e., at $u_r>M$), we have
\begin{equation}
W_n^{r}\approx \frac 1{2\sqrt{\pi }} \left( \left| I_n(u_{r})\right| ^2+\left|I_n(-u_{r})\right| ^2 \right) ,  \label{31}
\end{equation}
which describes a two-peak diffraction pattern, each of the peaks being slightly different from the initial Gaussian form.

\subsection{Summary}

Thus, we have presented a model of adiabatic asymmetric scattering of coherent superposition states of effective two-state atoms in the field of a standing wave. We have shown that the behavior of the atomic wave-packets at adiabatic diffraction may significantly differ from the diffraction at resonant scattering.

We have demonstrated that the discrete Gaussian wave packet at adiabatic diffraction undergoes highly asymmetric scattering ({\em refraction}) while the evolution of the same wave packet at resonant scattering can be characterized as a high-order narrowed scattering ({\em bi-refringence}). Hopefully, these peculiarities of the standing wave diffraction of atoms with Gaussian initial momentum distribution will be useful for atom interferometric and atom lithographic applications.
\\

\subsection{Acknowledgments}

This research has been conducted within the scope of the International Associated Laboratory (CNRS-France \& SCS-Armenia) IRMAS.
The research has received funding from the European Union Seventh Framework Programme under great agreement No. 295025 - IPERA. The work was supported by the Armenian State Committee of Science (Grant No. 13RB-052) and by the Fund for Basic Researches of Belarus (Grant No. F14ARM-021).

\end{document}